\documentclass[]{aastex631}
\usepackage{amsmath}

\begin{document}

\title{Power-law Indices of EUV Intensity Power Spectrum in Flaring Coronal Active Regions}

\correspondingauthor{Sihui Zhong}
\email{sihui.zhong@kuleuven.be}

\author[0000-0002-5606-0411]{Sihui Zhong}
\affiliation{Centre for mathematical Plasma Astrophysics, Department of Mathematics, KU Leuven, \\Celestijnenlaan 200B box 2400,
Leuven, BE-3001, Belgium}
\affiliation{Engineering Research Institute \lq\lq Ventspils International Radio Astronomy Centre (VIRAC)\rq\rq\ of Ventspils University of Applied Sciences,\\ Inzenieru iela 101, Ventspils, LV-3601, Latvia}

\author[0000-0002-0687-6172]{Dmitrii Y. Kolotkov}
\affiliation{Centre for Fusion, Space and Astrophysics, Department of Physics, University of Warwick,\\ Coventry CV4 7AL, UK}
\affiliation{Engineering Research Institute \lq\lq Ventspils International Radio Astronomy Centre (VIRAC)\rq\rq\ of Ventspils University of Applied Sciences,\\ Inzenieru iela 101, Ventspils, LV-3601, Latvia}

\author[0000-0001-6423-8286]{Valery M. Nakariakov}
\affiliation{Centre for Fusion, Space and Astrophysics, Department of Physics, University of Warwick,\\ Coventry CV4 7AL, UK}
\affiliation{Engineering Research Institute \lq\lq Ventspils International Radio Astronomy Centre (VIRAC)\rq\rq\ of Ventspils University of Applied Sciences,\\ Inzenieru iela 101, Ventspils, LV-3601, Latvia}

\begin{abstract}
Solar intensity power spectra are usually characterised by coloured noise, with the spectral energy following a segmented power-law function of frequency, $S(f)\propto f^{-\alpha}$, over different frequency ranges. 
Typically, the power-law index exceeds 1 in the low-frequency part ($\alpha_\mathrm{lf}$) and is around 0 at high frequencies ($\alpha_\mathrm{hf}$).
This work investigates the spatial and temporal evolution of the power-law indices of coronal EUV intensity power spectra in flare-hosting active regions.
The spatial distribution of the power-law index in the low-frequency domain ($\alpha_\mathrm{lf}$) closely mirrors EUV intensity images, indicating that $\alpha_\mathrm{lf}$ can reveal the dynamics of coronal plasma structures.
Temporally, $\alpha_\mathrm{lf}$ remains stable in quiescent active regions, but it exhibits significant variability before the flare onset.
Motivated by this behaviour, we analysed 14 flare events, quantifying the temporal variation of the indices $\alpha_\mathrm{lf}$ and $\alpha_\mathrm{hf}$ as potential flare precursors. In all flare events considered, notable deviations of $\alpha_\mathrm{lf}$ beyond a defined threshold consistently occurred at the flare site within a few minutes before the flare. In some cases, the change in the value of $\alpha_\mathrm{lf} - \alpha_\mathrm{hf}$ was detected within 30--90\, minutes before the flare.
This proof-of-concept study suggests that the temporal variation of the power-law indices in coronal EUV intensity power spectra could potentially serve as short-term precursors of solar flares, which needs to be validated on a larger flare sample.

\end{abstract}


\section{Introduction} \label{sec:intro}

Solar flares are sudden releases of magnetic energy in the solar atmosphere, manifesting as intense radiation across the entire electromagnetic spectrum. Flares can accelerate energetic particles, produce coronal mass ejections (CMEs), and trigger extreme space weather events that affect Earth and human activities \citep[see e.g.][for comprehensive reviews]{2011LRSP....8....6S, 2011SSRv..159...19F, 2017LRSP...14....2B, 2021ARA&A..59..445H}.

Current flare prediction methods largely rely on the morphology of the hosting active region (AR) or sunspot and empirical flare records, with photospheric magnetograms revealing the magnetic structure.
Flare productivity is known to correlate with the complexity of an AR's magnetic field \citep[e.g.,][]{2005ApJ...631..628M}, characterised by, for example, the McIntosh classification scheme \citep{1990SoPh..125..251M}. This is the basis of many operational forecasting models.
Additionally, ARs with higher apparent magnetic non-potentiality have been observed to be more flare-productive \citep{2007ApJ...655L.117S}. Based on this, a forecasting algorithm has been developed using the empirical relationship between the historical flare rates and a proxy for free magnetic energy \citep[][]{2011SpWea...9.4003F}.  
A more physically grounded forecasting approach would identify triggering mechanisms and predict when, where, and how large a flare might occur. However, flares can be triggered by various mechanisms \citep[e.g.,][]{2017LRSP...14....2B}, which cannot be fully explained by the existing standard flare model \citep{1964NASSP..50..451C,1966Natur.211..695S,1974SoPh...34..323H,1976SoPh...50...85K}.  
One example of physics-based predictive attempt is by \cite{2020Sci...369..587K}, who explored major flare prediction via runaway MHD instability initiated by magnetic reconnection near the polarity inversion line (PIL). Their model links the location and size of flares to the critical length of the primary reconnection and the available free magnetic energy. 
However, as shown in a comprehensive comparison by \cite{2019ApJS..243...36L}, none of the existing flare forecasting methods is universally reliable, reflecting the incomplete understanding of flare physics.
Given the growing volume of solar observational data and complexity of underlying mechanisms, artificial intelligence methods \citep[e.g.,][]{2025ApJ...985...53G}, including machine learning techniques \citep[e.g.,][]{2018SoPh..293...28F,2024ScChD..67.3727H,2024ApJS..274...38Z}, are increasingly considered as promising tools to assist flare forecasting. 

In parallel to model-driven methods, empirical forecasting approaches have long sought flare precursors, analogous to the earthquake foreshocks. Possible flare precursors include swarms of small-scale low-atmospheric energy releases \citep{2017NatAs...1E..85W}, strong gradients at PILs and shear magnetic flux \citep[see e.g.,][and references therein]{2019LRSP...16....3T}, nonthermal velocity increase \citep[e.g.,][]{2001ApJ...549L.245H}, and the appearance of hot (10--15~MK) plasma revealed by soft X-ray data \citep[e.g.,][referred to as \lq\lq Hot Onset Precursor Event (HOPE)\rq\rq]{2021MNRAS.501.1273H}. HOPE has been identified as a common phenomenon preceding the impulsive phase of a flare, and thus can be useful for short-term flare alerts \citep{2025SoPh..300....2H}. Recently, \citet{2020MNRAS.491.4435H} reported the positive correlation between flare waiting time and flare magnitude in two isolated AR, supporting the idea of magnetic energy build-up and release scenario. However, statistical analysis with larger samples did not confirm this correlation \citep{2020SoPh..295..132H}. 
Alternative empirical precursors include quasi-periodic pulsations in the observed light curves (QPPs; \citealt{2021SSRv..217...66Z}).
For example, QPPs in microwave emission before flares have been detected since the 1970s \citep[e.g.,][]{1973R&QE...16.1047B,2019Ge&Ae..59..822A}, with their spatial locations coinciding with the maximum microwave brightness during the subsequent flare \citep{2022Ge&Ae..62..895A}. 
In \citet{2022A&A...663A.149A}, 8-s pulsations seen in the microwave emission before the flare onset were linked with the dynamics of electric currents (as the main storage of free magnetic energy) and an oscillatory regime of the coalescence instability. However, the flare-predictive potential of such observations was not explored in the study.
Furthermore, \cite{2016ApJ...833..206T} found preflare very long-period QPPs in soft X-rays in about one-third of isolated flares, and interpreted them as oscillations of current-carrying loops as equivalent LRC electric circuits. Using RHESSI X-ray images, \cite{2022Ge&Ae..62..356Z} examined the spatial distribution of these preflare pulsations, finding that some originate in the flaring AR while others appear elsewhere.
However, none of these proposed precursors is universally present, which limits their predictive reliability. 

An important observable from the solar atmosphere is the intensity variation across multiple wavelengths, from low-frequency radio to hard X-rays and gamma-rays, detected with a fleet of spaceborne and ground-based telescopes.
Coronal observations with high resolution in both space and time are available mainly in the Extreme Ultraviolet (EUV) band. In particular, uninterrupted full-disk images are provided with the Atmospheric Imaging Assembly (AIA; \citealt{2012SoPh..275...17L}) EUV imager on board the Solar Dynamics Observatory (SDO; \citealt{2012SoPh..275....3P})  since 2010. 
The temporal fluctuations of solar intensity, at various atmospheric heights, often exhibit a power-law behaviour in the power spectra over a certain frequency range. Specifically, \cite{2007ApJ...657L.121M} identified flicker noise (with a $f^{-1}$ spectrum) at $f<0.6$~mHz in both the interplanetary medium and the photosphere and linked it with fluctuations in density and magnetic field. Intensity power spectra, $S(f)\propto f^{-\alpha}$, have been found to follow red noise ($f^{-2}$) at low frequencies and white noise ($f^0$) at high frequencies in the chromosphere \citep[e.g.,][]{2008ApJ...683L.207R}, corona \citep[e.g.,][]{2014A&A...563A...8A,2014A&A...568A..96G} and beyond \citep[e.g.,][]{2009ApJ...693.1022T,2009ApJ...706..238T}. \citet{2003A&A...409L..17G} established that fluctuations in the EUV/soft X-ray solar irradiance integrated over the full disk have a power-law spectrum.

The slope of these spectra, the power-law index $\alpha$, can vary locally depending on the plasma structure and can be considered as a quantitative measure of the corresponding dynamic processes developing in it. For example, \cite{2008ApJ...677L.137B} observed a two-segment power-law at low latitudes and a three-segment spectrum at high latitudes near 2 solar radii.
Studies by \citet{2015ApJ...798....1I,2016A&A...592A.153K} confirm that solar intensity fluctuations are composed of coloured noise, with power-law indices varying over different frequency ranges and atmospheric heights. For example, \citet{2016A&A...592A.153K} found that the power-law spectral index of the EUV emission intensity in 1600\,\AA\ channel of SDO/AIA varies from 0.86 to 1.33 in a quiet Sun region and above a sunspot, respectively.

Motivated by these findings, we propose that the value of the power-law index, $\alpha$, of intensity power spectra carries information about atmospheric plasma dynamics. We apply this idea to flare-hosting coronal active regions, focusing on the temporal and spatial variation of $\alpha$ before flares, which may indicate changing local plasma conditions and hence reflect the transition of the plasma configuration into a critical, flare-prone state.
In this work, we first construct spatial maps of the power-law indices to investigate their relationship with different plasma structures and their dynamics, using a full-disk image sequence.  
We then explore the temporal variability of the power-law index in the low-frequency domain, showing that it is quasi-static in non-flaring active regions but exhibits significant fluctuations in flaring areas. Inspired by this behaviour, we analyse 14 flare events to investigate preflare trends in the power-law indices in both the low and high-frequency parts of the spectrum, $\alpha_\mathrm{lf}$ and $\alpha_\mathrm{hf}$, respectively. For comparison, a quiescent active region is included. Our results suggest that the increase or decrease of $\alpha_\mathrm{lf}$ and the variability of $\alpha_\mathrm{lf} - \alpha_\mathrm{hf}$ happen before flares near the flare site.


\section{Methods} \label{sec:method}

\subsection{Observational data and processing}

In this work, to study the characteristics of power-law indices in the coronal intensity power spectra, we use EUV images taken by AIA/SDO. 
First, AIA full-disk images are considered to investigate the spatial dependence of the power-law indices in different coronal environments, such as quiet Sun regions, quiescent and flaring active regions, and off-limb background. In this case, a set of 171\,\AA\, images on 2023-03-29 from 08:00 to 16:00~UT with 12-s cadence is analysed.

Then, to study the time variation of the power-law indices before flares, we look at observations of certain active regions hosting flares. We only search for flare events which are \lq\lq isolated\rq\rq\ in time, in the SolarSoft FLARE Events Archive\footnote{\url{https://www.lmsal.com/solarsoft/latest_events_archive.html}}. The selection criteria are as follows. (1) To avoid sympathetic events that include the influence of the pre-existing flares. There should be no other flares at least 3 hours earlier than the start time of the candidates. (2) The region of interest (ROI) is on disk, best to be near the disk centre, to minimise the line-of-sight (LoS) integration effect. Among all six EUV passbands in AIA, we would select one or two wavelengths which show the signatures of a flare, such as the flashing brightening, flare ribbons, etc. Concerning the time interval of the image sequence, at least 2 hours before the flare event are included. The data analysed in this work are displayed in Table~\ref{table:1}.

After selection, with confirmation of the time range, field of view (FOV) of ROIs, and wavelength of the AIA images for each candidate, level 1 cut-out data are downloaded from the JSOC\footnote{\url{http://jsoc.stanford.edu/ajax/lookdata.html}}, before being processed with \texttt{aia\_prep.pro} for normalisation and upgrade to level 1.5 data. In addition, the de-rotation is done with \texttt{drot\_map.pro}. Then the data is ready for Fourier analysis as described in Sec.~\ref{sec:fourier}.

\begin{table}[!ht]
\centering
\begin{tabular}{ccccccccccc} 
 \hline
 Date & Start & Duration & AR & Flare class & Flare time  & Hale & Channel (\AA) & FoV centre & FoV size  \\  
 \hline\hline
 2015-01-04 & 18:15 & 4.5~h & 12253 & C1.1 & 21:21 -- 21:26 -- 21:35 & {$\beta\gamma\delta$} & 131 & [-20\arcsec,-75\arcsec] & 799\arcsec$\times580$\arcsec \\ 
 2013-12-31 & 13:35 & 3.8~h & 11938 & C2.1 & 16:42 -- 16:49 -- 16:55 & {$\alpha$} & 131 & [-290\arcsec,-180\arcsec] & 588\arcsec$\times439$\arcsec\\ 
 2023-04-26 & 21:00 & 4~h & 13288 & C2.7 & 23:57 -- 00:04 -- 00:09 & {$\beta$} & 131 & [-200\arcsec,-208\arcsec] & 600\arcsec$\times600$\arcsec\\ 
 2011-03-10 & 09:20 & 6~h & 11166 & C4.7 & 13:42 -- 13:45 -- 13:50 & {$\beta\gamma\delta$} & 171 & [-250\arcsec,269\arcsec] & 601\arcsec$\times601$\arcsec\\ 
 \hline\hline
 2014-01-07 & 01:10 & 3.1~h & 11946 & M1.0 & 03:49 -- 03:53 -- 03:56 & {$\beta\gamma$} & 131, 171 & [-150\arcsec,200\arcsec] & 400\arcsec$\times298$\arcsec \\ 
 2015-01-04 & 12:10 & 4~h & 12253 & M1.3 & 15:18 -- 15:36 -- 15:53 & {$\beta\gamma\delta$} & 131 & [-20\arcsec,0\arcsec] & 760\arcsec$\times600$\arcsec \\ 
 2023-03-29 & 08:00 & 8~h & 13262 & M1.3 & 13:45  -- 14:00 -- 14:15 & {$\alpha$} & 171 & [0\arcsec,0\arcsec] & Full disk \\ 
 2011-09-05 & 23:00 & 5~h & 11283 & M5.3 & 01:35 -- 01:50 -- 02:05 & {$\beta\delta$} & 171 & [80\arcsec,105\arcsec] & 721\arcsec$\times703$\arcsec\\ 
 2014-01-01 & 15:40 & 3.4~h & 11936 & M9.9 & 18:40 -- 18:52 -- 18:52 & {$\beta\gamma\delta$} & 193 & [670\arcsec,-260\arcsec] & 520\arcsec$\times520$\arcsec \\
 \hline\hline
  2012-07-12 & 12:00 & 6~h & 11520 & X1.4 & 15:37 -- 16:49 -- 17:30 & {$\beta\gamma\delta$} & 171 & [150\arcsec,-360\arcsec] & 902\arcsec$\times650$\arcsec\\
  2014-09-10 & 11:50 & 6.1~h & 12158 & X1.6 & 17:21 -- 17:45  & {$\beta\gamma\delta$} & 171 & [-140\arcsec,128\arcsec] & 600\arcsec$\times600$\arcsec \\
  2011-09-06 & 19:00 & 4.5~h & 11283 & X2.1 & 22:12 -- 22:20 -- 22:24 & {$\beta\gamma$} & 171 & [280\arcsec,185\arcsec] & 704\arcsec$\times717$\arcsec\\
  2014-10-24 & 18:00 & 4~h & 12192 & X3.1 & 21:07 -- 21:40 -- 21:41 & {$\beta\gamma\delta$} & 171 & [240\arcsec,-344\arcsec] & 900\arcsec$\times900$\arcsec\\
 \hline\hline
 {*2023-03-29} & {11:10} & {4.8~h} & -- & -- & -- & -- & {171} & {[199.5\arcsec,511.5\arcsec]} & {481\arcsec$\times481$\arcsec} \\
 \hline
\end{tabular}
\caption{Analysed image sequences with different flare classes. A quiescent AR (labelled with *) is included for comparison. All times in the table are in UT. The times in the column of flare time are the start time, peak time, and end time, respectively. For 2011-09-06, a C1.4 flare from 20:42 to 22:00~UT. For 2014-09-10, there is a C1.5 flare from 13:57-14:08~UT before the X-flare. For 2014-09-10 X1.6 flare, its peak time coincides with its end time. Here, \lq\lq Hale\rq\rq\ stands for the Hale (Mount Wilson) Class.}
\label{table:1}
\end{table}

\subsection{Fourier analysis}\label{sec:fourier}
\begin{figure}[ht!]
    \centering
    \includegraphics[width=\linewidth]{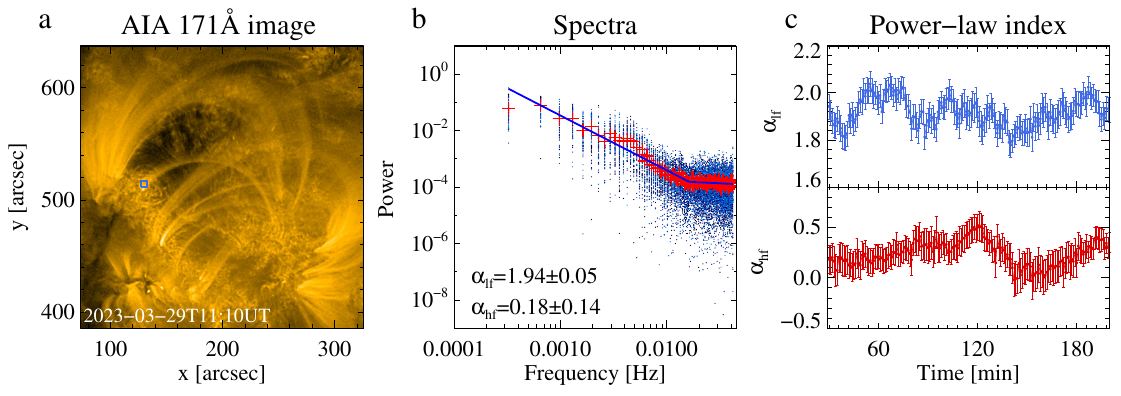}
    \caption{Demonstration of power-law index estimation in Fourier power spectra of time series of coronal EUV intensity. (a) An AIA 171\AA\, image showing an example of a region of interest. The blue box indicates the macropixel where the signals shown in panels b--d are extracted from. (b): Fourier power spectra of detrended signals in each pixel of the macropixel (blue dots). The red plus symbols represent the mean spectrum (see the red plus) and are then fitted by a broken power-law function (Eq.~\ref{eq:break}) with the power-law indices ($\alpha_\mathrm{lf}$ and $\alpha_\mathrm{hf}$) indicated in the panel. (c): Time variation of $\alpha_\mathrm{lf}$ and $\alpha_\mathrm{hf}$ from 11:30 to 13:00~UT. All times shown in this figure start from 11:00~UT.}
    \label{fig:demo}
\end{figure}

To estimate the power-law index from EUV imaging data, we analyse the temporal intensity variations as follows. The spatial domain of an EUV image sequence is divided into macropixels of size $a\times a$ pixels, with $a=10,32,64$ for different analysis scales. 
For each macropixel (see the blue box in Figure~\ref{fig:demo}(a)), we extract the time series of intensity from each pixel over a time window $\Delta T$. To minimise edge effects on the Fourier power spectrum, a linear spline interpolation connecting the start and end points is removed from each signal \citep{2017A&A...602A..47P,2018ApJ...858L...3K, 2019ApJS..244...44B}. The Fourier power spectrum of the detrended signal is then computed.
This process is repeated for all $a^2$ pixels in each macropixel. After that, the resulting spectra are averaged over the macropixel (see the average denoted by the red symbols in Figure~\ref{fig:demo}(b)). The uncertainty of the mean power in each frequency bin is determined by the spread of the power distribution from all $a^2$ pixels in each macropixel. Given that typical EUV intensity power spectra are characterised by distinctly different power-law indices in the high and low frequency parts \citep[e.g.,][]{2008ApJ...677L.137B,2015ApJ...798....1I,2016A&A...592A.153K}, the average power spectrum is best-fitted with a piecewise power-law function with a break point $f_0$:
\begin{align}\label{eq:break}
    S(f)= \begin{cases}
    c_1 f^{-\alpha_\mathrm{lf}} & f <f_{0}, \\
    c_2 f^{-\alpha_\mathrm{hf}} & f \geq f_{0},        
    \end{cases}
\end{align} 
\noindent where $\alpha_\mathrm{lf}/\alpha_\mathrm{hf}$ is the power-law index in low/high-frequency domain. To ensure continuity of the best-fit function given by Eq.~(\ref{eq:break}) at $f_0$, we use $f_0 = (c_2/c_1)^{1/{(\alpha_\mathrm{hf} - \alpha_\mathrm{lf})}}$. Hence, $f_0$ is not a fully independent parameter and will not be discussed in the following analysis. Our study focuses specifically on the two power-law indices. To characterise the spectrum with a broken power law, we fit the mean spectrum in log-log space, as shown in Figure~\ref{fig:demo}(b), where the relation reduces to a linear form. The fit was performed using a Levenberg--Marquardt least-squares method with \texttt{mpfitfun.pro}, and the parameter uncertainty corresponds to 1-$\sigma$ level. Here we ignore any local peak power above the noise spectrum, which is fitted by a lognormal function in \cite{2015ApJ...798....1I}, as we only care about the global spectral slopes $\alpha_\mathrm{lf}$ and $\alpha_\mathrm{hf}$. As shown in Figure~\ref{fig:demo}(b), the narrowband peak at around 4~mHz is much smaller relative to the full low-frequency range (0.9--10~mHz). It therefore does not significantly affect the robustness of the linear fit in the low-frequency domain. For example, Figure 1 of \citet{2016ApJ...833..284I} shows that the power-law indices estimated with and without the local power enhancement in the best-fitting model \citep[see models S0 and S1 in][]{2016ApJ...833..284I} differ by less than ~10\%, which is about our data analysis uncertainty. For a typical Fourier spectrum of EUV intensity in a coronal active region, see Figure~\ref{fig:demo}(b) where $\alpha_\mathrm{lf}>1$ and $\alpha_\mathrm{hf}\approx 0$. 
The fit error improves with longer $\Delta T$. To balance temporal resolution and error, we choose an optimal $\Delta T=30$~min. For the chosen time window, the uncertainty in $\alpha_\mathrm{lf}$ is typically below 0.1, i.e., less than 10\% of the typical values of $\alpha_\mathrm{lf}$ (also see Figure~\ref{fig:demo}(c)). The uncertainty in $\alpha_\mathrm{hf}$ is somewhat larger, since the fitted high-frequency segment is shorter. In some cases, it can be comparable to the absolute value of $\alpha_\mathrm{hf}$ given that the slope approaches zero. In addition, the model fidelity is justified by the reduced chi-squared value for the fits, with $\chi_{\nu}^2$ being within the reasonable range from 0.5 to 1.5.  For example, the representative fit in Figure~\ref{fig:demo}(b) gives $\chi_{\nu}^2=0.73$, indicating a reasonably good fit.
The above procedure is applied across all macropixels to generate spatial maps of power-law indices (see Fig.~\ref{fig:fulldisk}). By shifting the time window forward in 1-minute steps and repeating the analysis, we obtain the time evolution of these indices (Fig.~\ref{fig:demo}(c)). Here, the time of the parameters $\alpha_{\rm lf/hf}(t)$ denotes the end of the 30-minute time window, i.e., $[t-\Delta T, t]$. Typically, the values of $\alpha_\mathrm{lf}$ are found to be above 1 and $\alpha_\mathrm{hf}$ around 0, please also see statistics in Figure~\ref{fig:2dhist}. While $\alpha_{\rm hf}$ around zero is equivalent to a constant at high frequencies, keeping it as a free parameter is necessary for tracking its time evolution before the flare onset and for capturing its possible departures from zero (see Section~\ref{sec:time_variation}). Note that the errors of $\alpha_\mathrm{lf/hf}$ are quite constant in time, hence they are less important, compared to the time variation of $\alpha_\mathrm{lf/hf}$.


\section{Results} \label{sec:result}
The dynamics seen in the coronal EUV intensity signals are characterised by the power-law indices in the Fourier power spectra. Section~\ref{sec:space} illustrates the spatial distribution of the power-law indices across the full solar disk and their time evolution in several distinct regions of the Sun's atmosphere. In Section~\ref{sec:time_variation}, we focus on the time variation of the power-law indices before flares in several flaring ARs.


\subsection{Spatial dependence}\label{sec:space}

\begin{figure}
    \centering
    \includegraphics[width=0.95\linewidth]{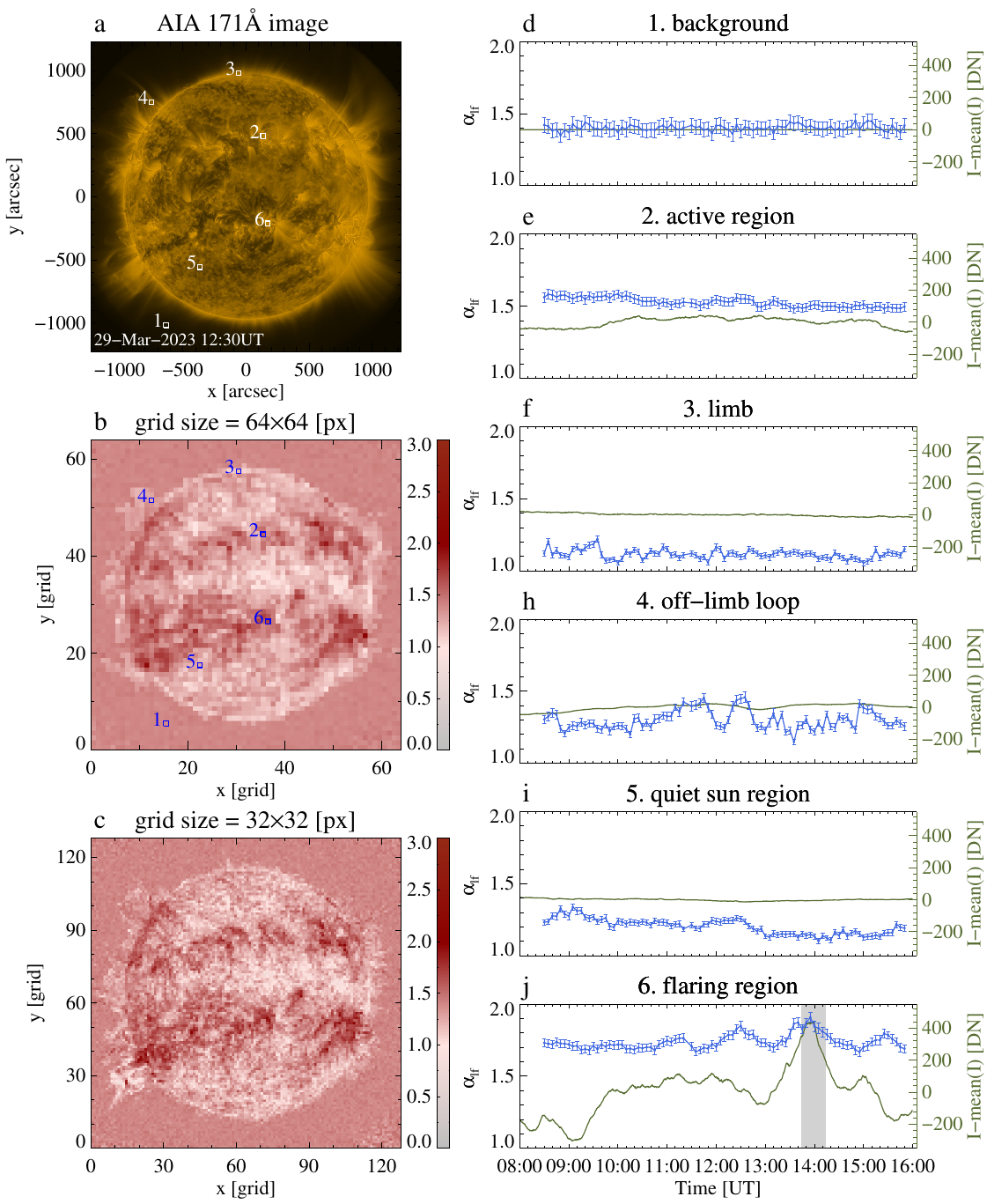}
    \caption{Power-law index map for a full-disk image sequence from 08:00~UT to 16:10~UT on 29th March 2023. (a) full-disk AIA 171\,\AA\, image with 6 regions of interest (ROIs, labelled 1--6) overplotted. (b--c): The map of $\alpha_\mathrm{lf}$ with different spatial resolution (64/32 pixels per macropixel in b/c). (d--e): Time evolution of intensity (green) and $\alpha_\mathrm{lf}$ (blue) in 6 selected ROIs. The grey region indicates the time interval of an M1.3 flare.}
    \label{fig:fulldisk}
\end{figure}

Using the methods described in Section~\ref{sec:method}, power-law index maps are computed for a sequence of AIA 171\,\AA\, full-disk images (Figure~\ref{fig:fulldisk}(a)) from 08:00 to 16:00~UT on 29th March 2023. In this analysis, the size of the macropixel is first taken to be $64\times64$ pixels and then refined to $32\times32$ pixels for a higher resolution, and the sliding time cadence of the maps is 5~min. 
At first glance, the map of $\alpha_\mathrm{lf}$ (Figure~\ref{fig:fulldisk}(b)) closely follows the structure of the AIA image (Figure~\ref{fig:fulldisk}(a)). This illustrates that the value of $\alpha_\mathrm{lf}$ is indeed related to the coronal environment. As displayed in Figure~\ref{fig:fulldisk}(b), $\alpha_\mathrm{lf}$ in regions including the limb, off-limb background, and the quiet sun region is around 1--1.5, which is consistent with the findings reported by \cite{2016A&A...592A.153K}. In contrast, in active regions, $\alpha_\mathrm{lf}$ is above 1.5 (see also panels e and j), and the solar activity belts in the north and south hemispheres are clearly visible in the obtained power-law index map.
The latter means that active regions of the solar corona contain more energy in the low-frequency part of the spectrum, and the characteristic time scales of their dynamics are longer than in quiet-Sun regions. This agrees well with a recent simulation by \citet{2017ApJ...834...10R} which demonstrated that the time scale of energy transport upwards in the quiet Sun and coronal arcade is about 20 and 50~min, respectively.
Upgrading the grid size of the power-law index map to $32\times 32$ pixels leads to a higher spatial resolution, which shows finer details. As seen in Figure~\ref{fig:fulldisk}(c), the higher values of $\alpha_\mathrm{lf}$ appear in the plasma structures, including loops, fans, and filaments.

The temporal behaviour of $\alpha_\mathrm{lf}$ varies across different coronal plasma structures. In the EUV emission detected from the limb, off-limb background and loops, quiet-Sun region, and quiescent active region, its variation is rather flat, see Figure~\ref{fig:fulldisk}(d--i). Here, the uncertainty in $\alpha_\mathrm{lf}$ is small and remains nearly constant, and thus has little impact on the subsequent discussion.
In contrast, visibly stronger variability in both the EUV intensity and $\alpha_\mathrm{lf}$ is seen in AR 13262, which hosts an M1.3-class flare from 13:45 to 14:15~UT.
As shown in Figure~\ref{fig:fulldisk}(j), $\alpha_\mathrm{lf}$ in a certain location in AR 13262 remains about 1.7 (practically constant) from 08:00 to 11:00~UT.
Then, at the time closer to the flare, the variability of $\alpha_\mathrm{lf}$ intensifies, reaching a maximum of about 1.9 near the time of the flare peak, potentially reflecting the change in plasma and/or magnetic field dynamics in the AR. The latter suggests the existence of an apparent correlation between the time variation of the power-law indices of coronal EUV intensity power spectra and the occurrence of flares, which we explore in Sections~\ref{sec:time_variation} and \ref{sec:occur-time}.



\subsection{Time variation before a flare} \label{sec:time_variation}
To analyse the time evolution of power-law indices before flares, 13 flaring ARs are studied. The 13 cases are displayed in Table~\ref{table:1}, including 4 C-class flares, 5 M-class flares, and 4 X-class flares. One quiescent AR on 29th March 2023 is also included for comparison. For each case, two sequences of power-law index maps are computed, one for $\alpha_\mathrm{lf}$ and the other for $\alpha_\mathrm{hf}$, with a grid size of $10\times10$ pixels and a time cadence of 1~min. Such high spatial and temporal resolution allows us to localise remarkable signatures in the evolution of $\alpha_\mathrm{lf}$ and $\alpha_\mathrm{hf}$ (which could carry information about potential flare triggers).

\begin{figure}
    \centering
    \includegraphics[width=\linewidth]{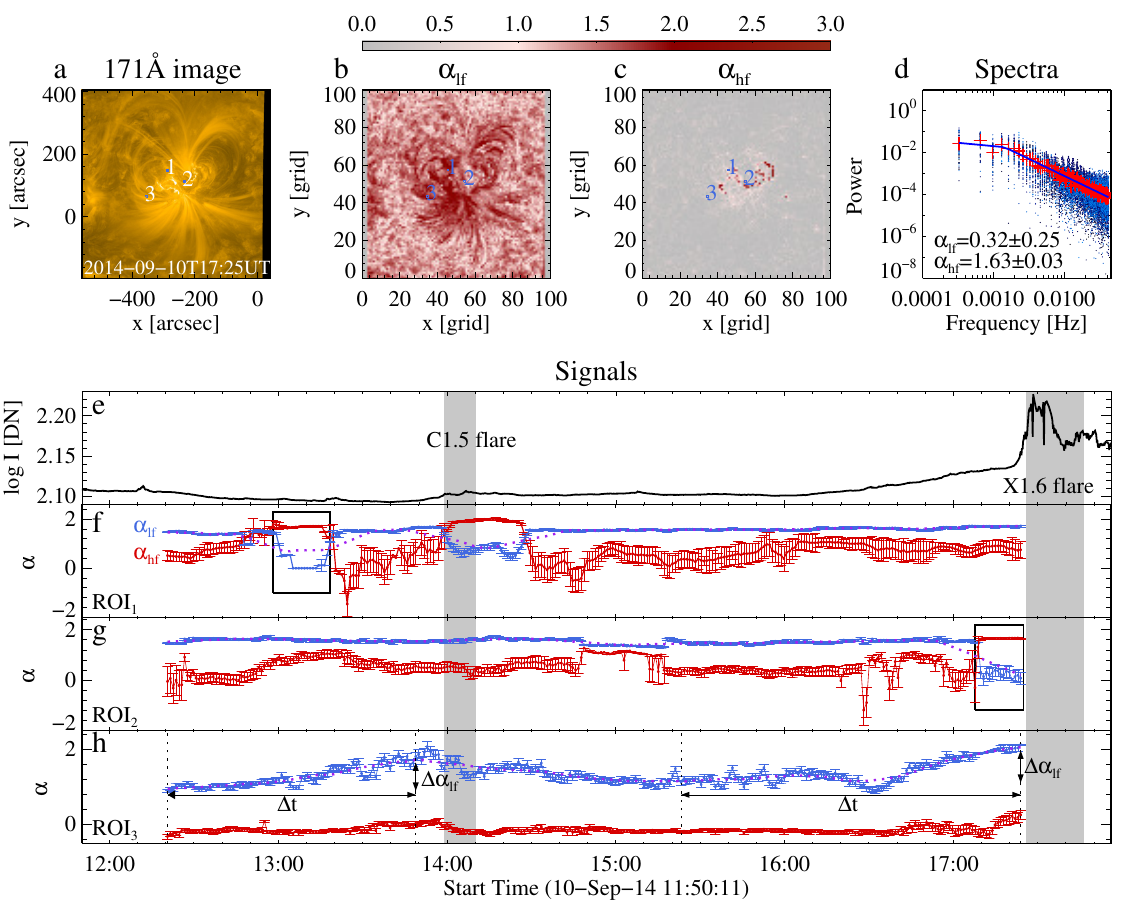}
    \caption{Time evolution of $\alpha_\mathrm{lf}$ and $\alpha_\mathrm{hf}$ in AR 12158 during 11:50 to 17:58~UT on 10th September 2014. (a): AIA 171\,\AA\ image showing the studied AR. The blue boxes labelled 1, 2 and 3 (ROI$_1$, ROI$_2$ and ROI$_3$) are three macropixels of interest, the first two that have the reversal of $\alpha_\mathrm{hf}$ and $\alpha_\mathrm{lf}$, i.e., $\alpha_\mathrm{hf}>1$ and $\alpha_\mathrm{lf}<1$, while ROI$_3$ sees the increase in $\alpha_\mathrm{hf}$ before flares. (b--c): Maps of $\alpha_\mathrm{lf}$ (b) and $\alpha_\mathrm{hf}$ (c) correspond to the image sequence. The grid size is $10\times10$\,pixels. The three blue boxes are the same as in (a). (d): abnormal Fourier spectra in ROI$_2$ with $\alpha_\mathrm{hf}>\alpha_\mathrm{lf}$. (e): time-varying averaged 171\,\AA\ intensity of AR 12158. The grey areas denote time intervals of C1.5 and X1.6 flares, respectively. (f--h): time series of power-law indexes in ROI$_1$, ROI$_2$ and ROI$_3$, respectively. The purple dotted curves are smoothed temporal signals of $\alpha_\mathrm{lf}$, which are used to calculate the actual difference and growth rate of $\alpha_\mathrm{lf}$. The black boxes mark the reversal of $\alpha_\mathrm{lf}$ and $\alpha_\mathrm{hf}$ before the flares. The vertical dotted lines denote the time of minimum/maximum smoothed $\alpha_\mathrm{lf}$ for two flares. The value and time difference between the minimum and maximum $\alpha_\mathrm{lf}$ are marked by $\Delta\alpha_\mathrm{lf}$ and $\Delta t$, respectively. An animation of panels (e) and (a--c) showing the temporal evolution of average intensity over the targeted active region, AIA 171\,\AA\ images, maps of the two power-law indices, and the three corresponding proxy maps derived from the power-law indices maps is available. The animation covers the period from 12:20 UT to 17:19 UT, with a real-time duration of 11 seconds.}
    \label{fig:reverse}
\end{figure}

An example case study on AR 12158 with two flares on 10th September 2014 is demonstrated in Figure~\ref{fig:reverse} and its accompanying animation 1. In this AR (Figure~\ref{fig:reverse}(a)), a C1.5 flare happens from 13:57 to 14:08~UT (see the first grey rectangle in Figure~\ref{fig:reverse}(e)); and 3.2 hours later, an X1.6 flare is produced (see the second grey rectangle in Figure~\ref{fig:reverse}(e)). Using 171\,\AA\ images covering this AR from 12:00 to 18:00~UT, map sequences of $\alpha_\mathrm{lf}$ and $\alpha_\mathrm{hf}$, respectively, are calculated. As shown in Figure~\ref{fig:reverse}(b--c), loop/fan structures are revealed by $\alpha_\mathrm{lf}\geq 2$, with $\alpha_\mathrm{hf}$ remaining around zero as expected. However, in certain areas, abnormal values of the power-law indices show up, specifically, $\alpha_\mathrm{lf}\approx0$ and $\alpha_\mathrm{hf}>1$ simultaneously in ROI$_1$ and ROI$_2$ (see the blue boxes). The corresponding spectra obtained in ROI$_2$ are shown in Figure~\ref{fig:reverse}(d), with $\alpha_\mathrm{lf}\approx0$ and $\alpha_\mathrm{hf}\approx2$, which is reverse to a typical Fourier spectrum of the coronal EUV emission intensity with $\alpha_\mathrm{lf}>1,\alpha_\mathrm{hf}\approx 0$. 
A possible reason to make $\alpha_\mathrm{hf}$ rise and $\alpha_\mathrm{lf}$ drop is the existence of a narrow spike of intensity (or a series of spikes) in the time series, perhaps caused by short-lived ($>1$~min) strong brightenings such as transients and jets that could be caused by episodes of magnetic reconnection. For this particular case, small-scale transients and brightening along threads occur intermittently at the core of the AR before the flares.
Looking at the time series of $\alpha_\mathrm{lf}$ (blue in Figure~\ref{fig:reverse}(f--g)) and $\alpha_\mathrm{hf}$ (red) in these two ROIs, such a reversal (outlined by black rectangle) occurs before the flare: one at around 13:00~UT that is 30~min before the C1.5 flare, another at around 17:10~UT just before the X1.6 flare. The duration of the reversal indicates the duration of the corresponding EUV brightenings.
This phenomenon appears in several macropixels around the flare site, see those that appear red in the fourth panel in the accompanying animation 1. More importantly, the fact that they appear in two separate macropixels may mean that the triggers of the two flares considered are rather local and spatially separated.
In addition, similar to the case discussed in Section~\ref{sec:space} (AR 13262), certain macropixels in the core of the AR 12158 exhibit a continuous increase of $\alpha_\mathrm{lf}$ before the flares. An example of such behaviour obtained in ROI$_3$ is displayed in Figure~\ref{fig:reverse}(h). Here, $\alpha_\mathrm{lf}$ is seen to increase from about 12:30 until 5~min before the C1.5 flare. Another increase in $\alpha_\mathrm{lf}$ occurs from about 16:30~UT until the onset of the X1.6 flare. At the same time, $\alpha_\mathrm{hf}$ in ROI$_3$ remains approximately constant, around 0. Therefore, the increase in $\alpha_\mathrm{lf}$ could be an additional signature of a flare trigger. 

\begin{figure}
    \centering
    \includegraphics[width=\linewidth]{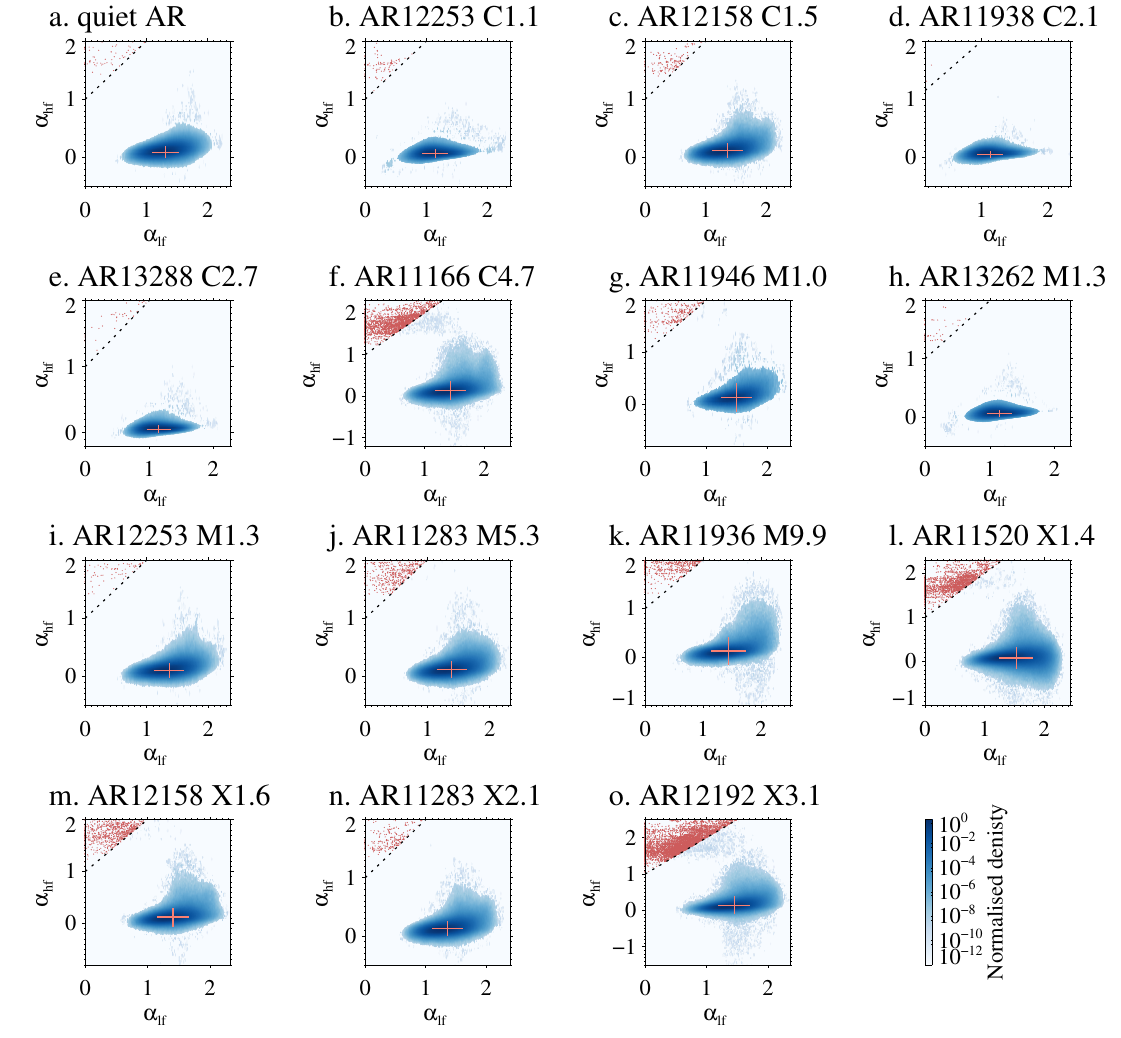}
    \caption{
    The 2D density distribution of $ \alpha_\mathrm{lf}$ vs. $\alpha_\mathrm{hf}$ in relative frequency for all analysed cases. The dotted lines indicate $ \alpha_\mathrm{hf}=1+\alpha_\mathrm{lf}$. The 2D density plots are normalised by their maxima. The location of the salmon horizontal/vertical lines indicates the mean of $\alpha_{lf}/\alpha_{hf}$, and the lengths indicate their standard deviations. 
    Note that the number of the reverse varies from hundreds to thousands across cases, and the total number of samples is around $10^6$, therefore, in some cases, the portion of $\alpha_\mathrm{lf}-\alpha_\mathrm{hf}\leq1$ takes less than 0.1\%, appearing as white in the contour plot. To make these minor portions visible, we overplotted them as red dots.
    }
    \label{fig:2dhist}
\end{figure}

Motivated by the example case and informed by the 14 analysed cases, we quantify the temporal and spatial evolution of $\alpha_{\rm lf}$ and $\alpha_{\rm hf}$ using the proxies defined below. As shown by the histograms of $\alpha_\mathrm{lf}$ and $\alpha_\mathrm{hf}$ and their 2D density distribution (Figure~\ref{fig:2dhist}) for all 15 cases, the typical value of $\alpha_\mathrm{lf}$ is above 1 and that of $\alpha_\mathrm{hf}$ is around 0. In most cases, $\alpha_\mathrm{lf}>\alpha_\mathrm{hf}$. However, abnormal $\alpha_\mathrm{lf} \rightarrow 0, \alpha_\mathrm{hf}>1$, occurs across events, see a notable example in Figure~\ref{fig:2dhist}(o). We refer to this as the reversal between $\alpha_\mathrm{lf}$ and $\alpha_\mathrm{hf}$, and we found that it is related to brightenings lasting several minutes. In this work, we quantify the minimal difference between two indices in time by the quantity $\min(\alpha_\mathrm{lf}-\alpha_\mathrm{hf})$ and define the reversal between $\alpha_\mathrm{lf}$ and $\alpha_\mathrm{hf}$ by the condition $\min(\alpha_\mathrm{lf}-\alpha_\mathrm{hf})\leq-1$. For the example case on 10th September 2014, the time sequence of $(\alpha_\mathrm{lf}-\alpha_\mathrm{hf})$ map is displayed in the fifth panel of animation 1, the minimal values in each macropixels in time are recorded to reveal the existence of the reversal between $\alpha_\mathrm{lf}$ and $\alpha_\mathrm{hf}$.

\begin{figure}
    \centering
    \includegraphics[width=\linewidth]{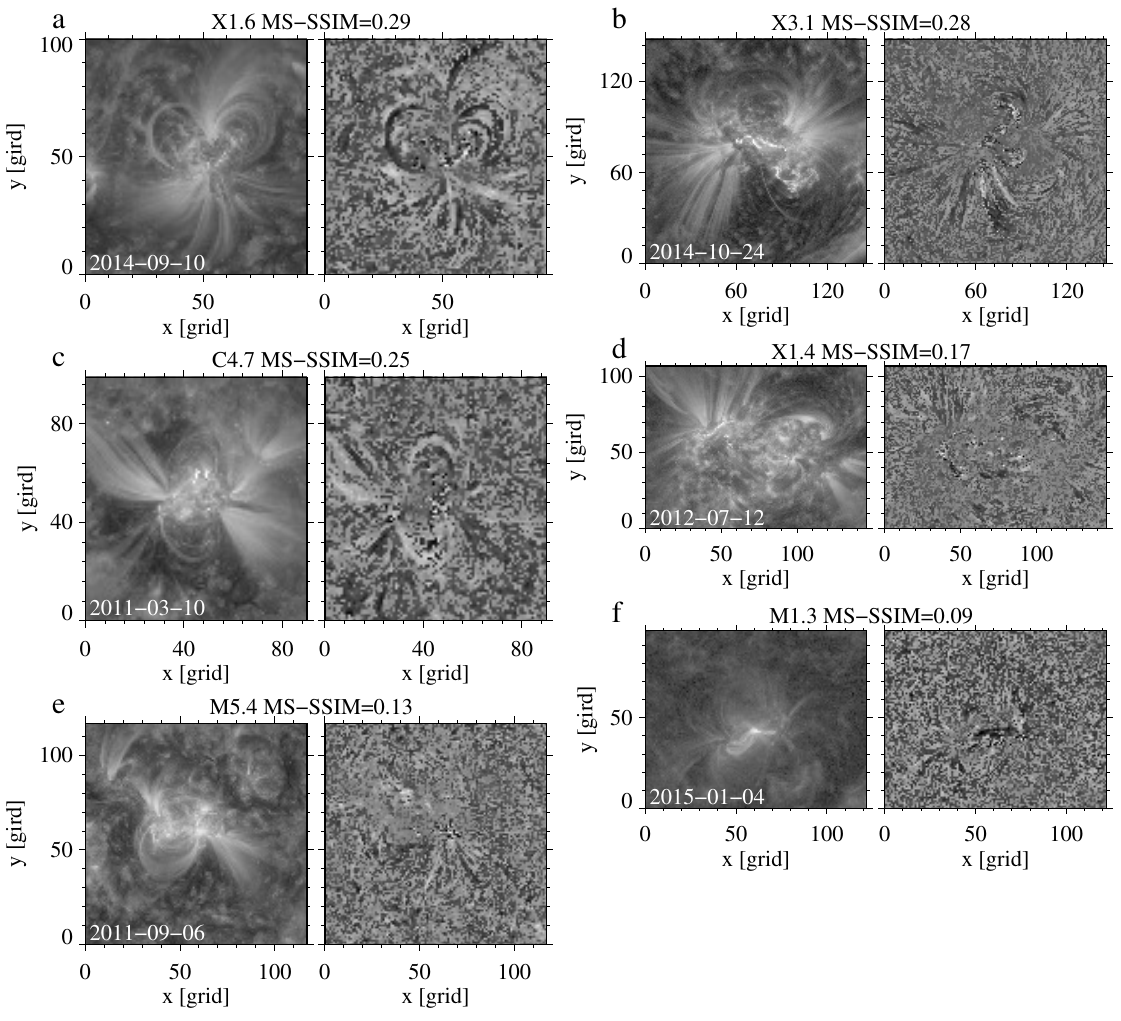}
    \caption{
    The multi-scale structural similarity measure (MS-SSIM) between the AIA image (left in each pair) and $\Delta\alpha_\mathrm{lf}$ map (right in each pair) for selected cases. The AIA images are degraded to the same resolution as the proxy maps.
    }
    \label{fig:msssim}
\end{figure}

To quantify the growth of $\alpha_\mathrm{lf}$, its trend is first obtained by smoothing the time series over a window of 30~min, see the purple dotted curves in Figure~\ref{fig:reverse}(f--h) as examples, to exclude the instantaneous minor fluctuations. 
Then the difference between the minimum and maximum of the trend, $\Delta\alpha_\mathrm{lf}$, is calculated, with a positive/negative sign indicating an increase/decrease. Further, the growth rate of $\alpha_\mathrm{lf}$ can be measured as $\Delta\alpha_\mathrm{lf}/\Delta t$
, where $\Delta t$ is the time difference between the minimum and maximum of $\alpha_\mathrm{lf}$. An example map sequence of $\Delta \alpha_\mathrm{lf}$ and $\Delta\alpha_\mathrm{lf}/\Delta t$ is displayed in the sixth and seventh panel of animation 1 to show how they respond to the 171\,\AA\ emission evolution before flares. In this animation, regions with high absolute proxy values are visually associated with coronal loops originating from the AR core. We use the term ``similarity" to denote this qualitative resemblance between the structures highlighted by the proxies and those apparent in the high-contrast intensity images. To enhance the visibility of this apparent similarity, the AIA image and the proxy map are shown in grayscale in Figure~\ref{fig:msssim}. Quantitatively, the multi-scale structural similarity measure (MS-SSIM, \citealt{1292216}) is calculated to evaluate the apparent similarity between the $\Delta\alpha_{\rm lf}$ maps and the intensity images. The MS-SSIM value ranges from 0 to 1, where 0 indicates no similarity, and 1 represents identical maps. This metric combines high-resolution intensity information with texture and contrast features across multiple spatial scales. And it is implemented via \texttt{sewar.full\_ref.msssim}\textcolor{red}{\footnote{\url{https://quality.nfdi4ing.de/en/latest/image_quality/MultiScale_Structural_Similarity.html}}} in Python.
For example, our tests showed that the MS-SSIM between the two AIA images, one original and the other contaminated with random noise with amplitudes ranging from 10\%, 20\% and 50\% of maximum intensity, is 0.92, 0.80, and 0.47, respectively. Six representative examples with different MS-SSIM values are shown in Figure~\ref{fig:msssim}. The resulting MS-SSIM values between the AIA images and the proxy maps are generally low (typically below 0.5, and in some cases as low as 0.1), reflecting the fact that the AIA images contain finer structural details, while the proxy maps are comparatively noisier. The MS-SSIM value shows a consistent correspondence with our visual assessment, being higher in panels (a)–(c) (about 0.3) where the similarity is visually better, and lower in panels (d)–(f) (about 0.1) where almost no similarity is observed. 

In the following analysis, we consider these three proxies for 14 flare case studies to evaluate their behaviour before flare onsets: (1) $\min(\alpha_\mathrm{lf}-\alpha_\mathrm{hf})$; (2) $\Delta \alpha_\mathrm{lf}$; and (3) $\Delta\alpha_\mathrm{lf}/\Delta t$.
Figure~\ref{fig:summary} shows the AIA images and corresponding maps of those three proxies in 4 representative cases. In each map, macropixels/locations where three proxies meet the following criteria are outlined by red contours: $\min(\alpha_\mathrm{lf}-\alpha_\mathrm{hf})\leq-1$, $\Delta\alpha_\mathrm{lf}=\pm0.6\max(\Delta\alpha_\mathrm{lf})$, $\Delta\alpha_\mathrm{lf}/\Delta t=\pm0.7\max(\Delta\alpha_\mathrm{lf}/\Delta t)$. The threshold of $\min(\alpha_\mathrm{lf}-\alpha_\mathrm{hf})$ is determined by the reversal between typical values of two indices (typically staying around 1 and 0). To determine an appropriate threshold for identifying high-value regions in the $\Delta\alpha_\mathrm{lf}$ maps, we adopted an empirical approach guided by morphological correspondence with EUV observations. Specifically, we increased the threshold value progressively from zero, expressed as a percentage of the maximum of the proxy, until the regions above the threshold exhibited (visually) clear spatial connectivity with the plasma structures identified in EUV images. Across all examined cases, a threshold of 60\% consistently produced the best correspondence, ensuring that the highlighted regions were both spatially coherent and physically meaningful. For $\Delta\alpha_\mathrm{lf}/\Delta t$ maps where the high-value region does not always align as clearly with EUV structures, the selection criterion was instead to ensure that the distribution of high-value macropixels was less random and more spatially organised. Applying the same procedure, we found that a 70\% threshold provided the most coherent distribution in these cases. We therefore adopted the 60\% and 70\% level as the working threshold of $\Delta\alpha_\mathrm{lf})$ and $\Delta\alpha_\mathrm{lf}/\Delta t$ for our analysis. 

In Figure~\ref{fig:summary}, the first row (a) is for a quiescent AR (no NOAA number is given, as it is a region between two big ARs).
For the quiescent AR, $\min(\alpha_\mathrm{lf}-\alpha_\mathrm{hf})\leq-1$ appear near a magnetic footpoint, and the maps of $\Delta\alpha_\mathrm{lf}$ and $\Delta\alpha_\mathrm{lf}/\Delta t$ appear random and speckled, see Figure~\ref{fig:summary}(a3)--(a4). The appearance of the reversal of $\alpha_\mathrm{lf}$ and $\alpha_\mathrm{hf}$ in panel (a2) correspond to small-size intermittent jets or elongated brightness from 11:50 to 14:10~UT, and a small-scale magnetic reconnection during 15:50 to 16:00~UT. Moreover, macropixels with high values of $\min(\alpha_\mathrm{lf}-\alpha_\mathrm{hf})\leq -1$ or $\Delta\alpha_\mathrm{lf}$ are isolated, not as collective as those in flaring active regions.
The remaining rows (b)-(d) are for flaring ARs, with each representing a C/M/X class flare. On the AIA images, the flaring structures that exhibit a strong intensity enhancement are outlined by red contours.
In contrast to a non-flaring active region (row (a)), in flaring cases, the maps of $\min(\alpha_\mathrm{lf}-\alpha_\mathrm{hf})$ and $\Delta\alpha_\mathrm{lf}$ contain more distinct features and, in general, show similarity with the AIA images. 
Unlike in the non-flaring region (row (a)), these outlined locations appear coherently/collectively and overlap well with the flaring site, as shown in Figure~\ref{fig:summary}(b1)--(d4). Thus, the collective appearance of the three proxies above a certain threshold may indicate that the system has entered a state prone to flare production.
In general, the maps of all three considered combinations of $\alpha_\mathrm{lf}$, $\alpha_\mathrm{hf}$, and $\Delta t$ share a similarity in structure\footnote{Note that the location of $\min(\alpha_\mathrm{lf}-\alpha_\mathrm{hf})\leq -1$ may correspond to positive (in green colour) or negative (pink) $\Delta \alpha_\mathrm{lf}$, depending on whether the dip of $\alpha_\mathrm{lf}$ occur after/before its maximum.}. 
However, the growth rate $\Delta\alpha_\mathrm{lf}/\Delta t$ maps are more poorly structured: their high-value areas appear to be less connected to actual plasma structures.
The location consistency between the flare site and the proxy thresholds implies that these proxies could be signatures of a flare trigger.
Corresponding animations showing the time evolution of the proxy maps are provided online\footnote{\url{10.5281/zenodo.16100847}.}. 
In general, the reversal between two indices, represented by $\alpha_\mathrm{lf}-\alpha_\mathrm{hf} \leq -1$ (see those in dark red colour on the map of $\alpha_\mathrm{lf}-\alpha_\mathrm{hf}$), occurs all the time, but in some cases it appears coherently (i.e., spatially associated with particular brightening structures) and frequently (i.e., repeatedly within the same temporal interval) before the flare. For the other two proxy maps, long before a flare, no high values exceeding the selected thresholds appear, and the spatial distribution resembles a random \lq\lq salt and pepper\rq\rq\ pattern. As the flare onset approaches, values exceeding the thresholds begin to appear collectively in all three maps. 

However, we did not manage to detect a correlation between the value of $\Delta\alpha_\mathrm{lf}$, $\Delta\alpha_\mathrm{lf}/\Delta t$ and the flare class. 
Figure~\ref{fig:correlation}(a--b) displays the distribution of $\Delta \alpha_\mathrm{lf}$ and $\Delta\alpha_\mathrm{lf}/\Delta t$ for each case. The peak and maximum values of the index as a function of flare class, as shown in panels (c--f), do not exhibit a clear correlation.
In addition, the spatial coherence of proxy maps and the similarity between the AIA image and proxy maps are considered.
For spatial coherence, it is qualitatively evaluated by visual checking if the high-value region is spatially distributed along a plasma structure, such as loops. A higher coherence can be seen in every class of flare, e.g., the 10th March 2011 C4.7 event and the 10th September 2014 X1.6 event. The apparent similarity is also determined by visual inspection, by comparing the resemblance between the high-intensity-contrast structures in the EUV image and high-value structures in the proxy map. This can also be quantified by the MS-SSIM (Figure~\ref{fig:msssim}). Comparable ranges of MS-SSIM values are found across all flare classes, indicating that the degree of apparent similarity is not correlated with flare magnitude. 



\subsection{Occurrence time}\label{sec:occur-time}
In the above, the spatial distributions of the three proxies $\min(\alpha_\mathrm{lf}-\alpha_\mathrm{hf})$, $\Delta\alpha_\mathrm{lf}$, and $\Delta\alpha_\mathrm{lf}/\Delta t$ are demonstrated. In this section, we investigate the occurrence time of the two of them ($\min(\alpha_\mathrm{lf}-\alpha_\mathrm{hf})$ and $\Delta\alpha_\mathrm{lf}$) exceeding a certain threshold relative to the flare time. Occurrence time refers to the end of the 30-minute interval over which intensity is analysed.
For each case, the macropixels with proxy values above a selected threshold are chosen, and their corresponding times are collected to check if they share a tendency in occurrence. For $\min(\alpha_\mathrm{lf}-\alpha_\mathrm{hf})$, the values $<-1$ are wanted; for $\Delta \alpha_\mathrm{lf}$, absolute values $>0.6\max(\Delta \alpha_\mathrm{lf})$ (which is around 1 in most of cases, the lowest is 0.7) are selected. For positive/negative $\Delta \alpha_\mathrm{lf}$, the time of maximum/mimumum $\alpha_\mathrm{lf}$ is considered.
Figure~\ref{fig:time} shows the time when the proxy thresholds happen relative to the flare time for all 14 cases. The C1.5 flare event, which occurred 3.2 hours before the X1.6 flare on 10th September, 2019 (see Section \ref{sec:time_variation}), is also included and treated separately in the time domain.

For reversal between $\alpha_\mathrm{lf}$ and $\alpha_\mathrm{hf}$, i.e., $\alpha_\mathrm{lf}-\alpha_\mathrm{hf}<-1$, it occurs when there could be transients, such as reconnection jets and brightenings. If there are no eruptions, the transients should possibly happen randomly and sporadically, leading to a uniform distribution of their temporal occurrence. When an energy release is triggered, or the configuration reaches the threshold of instability, such as in the form of flares, such transients/ejections could happen collectively and more frequently before the flare, leading to a peak distribution of our proxy $\alpha_\mathrm{lf}-\alpha_\mathrm{hf}<-1$ over a certain time interval. 
Among the 14 flare events considered, we could distinguish three types of behaviour in the observed distributions of $\alpha_\mathrm{lf}-\alpha_\mathrm{hf}<-1$: (1) with distinct peaks before flares (C1.1, C2.1, C2.7, M1.0, and both M1.3 flares); (2) occurs anytime with distinct peaks at a certain time before flares (C1.5, C4.7, M9.9, X1.4, and X1.6 flares); and (3) occurs anytime without distinct peaks and/or clear tendency (M5.3, X2.1 and X3.1 flares). For example:
\begin{itemize}
    \item In the C1.1 event on 4th January 2015, $\min(\alpha_\mathrm{lf}-\alpha_\mathrm{hf})\leq -1$ happens at around 30 min before the flare (at 21:21~UT). They correspond to a Y-shaped jet from 20:34 to 20:37~UT and an elongated jet from 20:37 to 20:42~UT along a fan anchored at a sunspot.
    
    \item In the C1.5 flare case on 10th September 2014, $\min(\alpha_\mathrm{lf}-\alpha_\mathrm{hf})\leq -1$ occurs within 90~min to the flare, with a distinct peak around 15~min. 
    From the sequence start, brightnings appear occasionally in both space and time, resulting in a uniform distribution of $\min(\alpha_\mathrm{lf}-\alpha_\mathrm{hf})\leq -1$ for this event at 100--35~min to the flare. When approaching the flare time, dotted and thread-like brightenings appear frequently and collectively. In particular, a thread-like brightening occurs during 13:16 to 13:22~UT, 40 min before the flare onset at 13:57~UT, causing the peak distribution at around 15 min.

    \item In the M1.0 flare case on 7th October 2014, the three peaks correspond to three collective occurrences of brightenings. For example, the first peak represents the intermittent magnetic reconnection (and hence brightenings) that occurs from 03:33 until the flare erupts at 03:49~UT, and the second peak corresponds to the jet along a fan from 02:15 to 02:20 UT and magnetic reconnection from 02:16 to 02:29 UT.
\end{itemize}

And a few less obvious examples:
\begin{itemize}
    \item In the C4.7 case, the peak time is broad and appears at around 2 hours away from the flare, suggesting a rather weak connection to the flare.

    \item In the cases of M5.3 and X3.1 flare, $\min(\alpha_\mathrm{lf}-\alpha_\mathrm{hf})\leq -1$ happens anytime before the flare, and the obtained distributions are rather flat. This could mean that the trigger process starts earlier and builds up longer to excite a final strong magnetic release. 

    \item In the X2.1 event, the reversal between two power-law indices happens rather sporadically without a clear pattern.
\end{itemize}

In summary, given that the condition $\min(\alpha_\mathrm{lf}-\alpha_\mathrm{hf})\leq -1$ is generally related to brightenings in reconnection-indicative morphologies such as Y-shaped, jet-like, and dot-like structures, $\min(\alpha_\mathrm{lf}-\alpha_\mathrm{hf})\leq -1$ could be used as an indicator of small-scale magnetic reconnection, maybe not necessarily related to a flare. Indeed, the supplementary movies of the $\min(\alpha_\mathrm{lf} - \alpha_\mathrm{hf})$ maps show that typically $\min(\alpha_\mathrm{lf} - \alpha_\mathrm{hf}) < -1$  occurs at various times before a flare, and in rather random locations, i.e. non-collectively. However, its collective occurrence could be the signature of a flare trigger, as it suggests a strong and larger-scale magnetic reconnection, which is more likely to excite a flare.
In each flare event, the spatial and temporal behaviour of $\min(\alpha_\mathrm{lf}-\alpha_\mathrm{hf})\leq -1$ varies, possibly reflecting that individual flares may have different triggering mechanisms.
An alternative explanation for intensity enhancement is an increased amount of plasma along the LoS. Considering the typical duration of these brightenings is only several minutes, the trigger for the mass increase must be transient in nature.

Considering the occurrence time of large values of $|\Delta\alpha_\mathrm{lf}|$, ($\geq 0.6\max(\Delta\alpha_\mathrm{lf})$, see Figure~\ref{fig:time}(b)), they are found to appear mainly in the last minute to the flare. For clarification, this timing is based on a 1-minute cadence, determined by the 1-minute sliding time of the 30-min window, set in Section~\ref{sec:fourier}. As suggested by the map of $\Delta\alpha_\mathrm{lf}$ where both positive and negative maxima match the flare sites (Figure~\ref{fig:summary}), we consider absolute values of $\Delta\alpha_\mathrm{lf}$ here. In general, the number of macropixels for $\Delta\alpha_\mathrm{lf}\geq 0.6\max(\Delta\alpha_\mathrm{lf})$ and $\Delta\alpha_\mathrm{lf}\leq -0.6\max(\Delta\alpha_\mathrm{lf})$ are more or less the same\footnote{Except C1.1 event on 4th January 2015 with 41 vs. 168 macropixels with positive/negative values of $\Delta\alpha_\mathrm{lf}$. The histogram of occurrence time of positive values only, i.e., $\Delta\alpha_\mathrm{lf}\geq 0.6\max(\Delta\alpha_\mathrm{lf})$, is similar to that shown in Figure~\ref{fig:time}(b).}. This common behaviour implies a high correlation between $\Delta\alpha_\mathrm{lf}$ and the flare onset, and perhaps a very short-term predictive potential.
In particular, in the cases in X1.6/C1.5 on 10th September 2014 and M1.3 on 29th March 2023 (labeled by 2 in Figure~\ref{fig:time}), their occurrence time of $\Delta\alpha_\mathrm{lf}$ peak in the last minute are consistent with the observed continuous increase of $\alpha_\mathrm{lf}$ till the flare, as illustrated in Figure~\ref{fig:fulldisk}(h) and Figure~\ref{fig:reverse}(h) respectively. 


To summarise, both $\min(\alpha_\mathrm{lf}-\alpha_\mathrm{hf})\leq -1$ and $\Delta\alpha_\mathrm{lf}$ show spatial coherence and are consistent with the flare site. The temporal occurrence of $\min(\alpha_\mathrm{lf}-\alpha_\mathrm{hf})\leq -1$ in 14 flare events considered could be characterised by several categories, which may indicate different flaring mechanisms. 
The occurrence time of the threshold of $\Delta\alpha_\mathrm{lf}$ displays a robust temporal connection to the flare, suggesting its potential role as a short-term flare precursor. 
These can be confirmed in the supplementary movies, which show the time evolution of each proxy map.
In most cases, the number of macropixels with $\Delta\alpha_\mathrm{lf}\geq0.6\max(\Delta\alpha_\mathrm{lf})$ peaks at the last minute before the flare.
Moreover, the peak relative frequency of occurrence of $\min(\alpha_\mathrm{lf}-\alpha_\mathrm{hf})\leq -1$ shows a decreasing tendency as flare magnitude grows (see Figure~\ref{fig:cc_freq}(a)). Indeed, the Pearson correlation coefficient between the peak occurrence frequency and flare magnitude is found to be about $-0.5$, and the best-fitting linear model gives a negative slope of $-0.01$. For $|\Delta\alpha_\mathrm{lf}|$, ($\geq 0.6\max(\Delta\alpha_\mathrm{lf})$, no meaningful tendency is found.

\section{Summary and Discussion}\label{sec:discussion}

Fourier power spectra of coronal EUV intensity time series are approximated using a segmented power-law behaviour with two indices ($\alpha_\mathrm{lf}$ and $\alpha_\mathrm{hf}$) characterising the slope in the low- and high-frequency domains, respectively. A typical value of $\alpha_\mathrm{lf}$ is around 1--3, and that of $\alpha_\mathrm{hf}$ is around zero.
These power-law indices, especially $\alpha_\mathrm{lf}$, are associated with dynamic processes in the coronal plasma.
In this work, we demonstrate that $\alpha_\mathrm{lf}$ has different values in different regions in the corona, with higher values in more dynamic plasma structures such as coronal loops and fans. The spatial maps of $\alpha_\mathrm{lf}$  resemble EUV images. Furthermore, the index $\alpha_\mathrm{lf}$ does not change significantly in time in quiescent regions \citep[cf. observations of quiescent diffusion regions by][]{2023A&A...678A.188G}, while it shows strong variability, such as an increase, before a flare in a flaring region.
In addition, reversals between $\alpha_\mathrm{lf}$ and $\alpha_\mathrm{hf}$ are detected before the C1.5 and X1.6-class flares in AR 12158 on 10th September 2014, which was attributed to the appearance of small-scale brightenings and other transients.

To investigate the general trend of time variation of power-law indices before flares, 14 ARs with imaging data in at least 2 hours before the flare are studied by computing sequences of spatial maps of three proxies: $\min(\alpha_\mathrm{lf}-\alpha_\mathrm{hf})$, $\Delta \alpha_\mathrm{lf}$ and $\Delta\alpha_\mathrm{lf}/\Delta t$. Here, $\min(\alpha_\mathrm{lf}-\alpha_\mathrm{hf})<-1$ represent the reversal between $\alpha_\mathrm{lf}$ and $\alpha_\mathrm{hf}$, $\Delta \alpha_\mathrm{lf}$ characterises the increase/decrease of $\alpha_\mathrm{lf}$, and $\Delta\alpha_\mathrm{lf}/\Delta t$ show the growth rate of $\alpha_\mathrm{lf}$.
All cases show visual similarity between the maps of $\Delta\alpha_\mathrm{lf}$ or $\min(\alpha_\mathrm{lf}-\alpha_\mathrm{hf})$, and the AIA image, except for the non-flaring AR. 
The map of the growth rate $\Delta\alpha_\mathrm{lf}/\Delta t$ is less structured because the large values of $\Delta\alpha_\mathrm{lf}$ could be averaged over time.
Additionally, in all flaring cases, the locations with values exceeding selected thresholds (e.g., $\min(\alpha_\mathrm{lf}-\alpha_\mathrm{hf})<-1$, $|\Delta \alpha_\mathrm{lf}|\geq0.6\max(\Delta \alpha_\mathrm{lf})$) appear to show coherent behaviour, and they overlap with the flare site. The spatial distribution of growth rates $\Delta\alpha_\mathrm{lf}/\Delta t$ greater than its 70\% maximum does not exhibit clear structure and does not seem to spatially resemble the structure of EUV intensity images, though a portion of them agree with the flare site.
The location consistency between the maps of $\min(\alpha_\mathrm{lf}-\alpha_\mathrm{hf})$ or $\Delta\alpha_\mathrm{lf}$ and the flare site suggests they could act as potential precursors of a flare.

Naively, one might expect the magnitude of a possible triggering process to correlate with the flare magnitude. However, we could not observe a meaningful correlation between these (their values, spatial coherence, and similarity with AIA images) and the flare magnitude. In other words, the energetics of a potential upcoming flare cannot be inferred from the observed behaviour of our proxies. A preliminary indication of a negative correlation is seen between the relative occurrence frequency of power-law index reversals and the flare class, as suggested by a best-fit line with a negative slope in Figure~\ref{fig:cc_freq}(a). However, the Pearson correlation coefficient was found to be just about $-0.5$, and needs validation on a larger flare sample. The apparent lack of correlation may also reflect a threshold-triggered nature of the flare process, in which the flare magnitude becomes independent of the trigger amplitude once the flare onset occurs.

The occurrence time of $\min(\alpha_\mathrm{lf}-\alpha_\mathrm{hf})<-1$ and $|\Delta\alpha_\mathrm{lf}|\geq 0.6\max(\Delta\alpha_\mathrm{lf})$ in 14 flaring cases is investigated to check their potential as a flare precursor. In 9 of 14 events, the occurrence time histogram for $\min(\alpha_\mathrm{lf}-\alpha_\mathrm{hf})<-1$ shows distinct peaks at certain times. The peak times vary from tens of minutes to 2.5 hours before flares. The peak distribution represents the collective and more frequent occurrence of $\min(\alpha_\mathrm{lf}-\alpha_\mathrm{hf})<-1$. The remaining events exhibit rather flat distributions (sometimes with broad peaks), indicating that $\min(\alpha_\mathrm{lf}-\alpha_\mathrm{hf})<-1$ occurs most of the time before the flare, but at some certain time it appears collectively.
In turn, $\Delta\alpha_\mathrm{lf}$ shows a shorter-term temporal connection to the flare.
For all events, $|\Delta\alpha_\mathrm{lf}| \geq 0.6\max(\Delta \alpha_\mathrm{lf})$ mainly occurs within 1~min before the flare. 
These findings suggest that observations of $\min(\alpha_\mathrm{lf}-\alpha_\mathrm{hf})<-1$ appearing collectively in an AR may indicate that a flare will occur within the next 30~min to 2~hours, whilst a high value of $|\Delta\alpha_\mathrm{lf}|$ could be a signature of a flare to occur within the next few minutes.

It has been empirically established that the spectra of EUV intensity variation in the high-frequency domain are usually white-noise-like \citep[e.g.,][]{2015ApJ...798....1I,2016A&A...592A.153K}, i.e., the high-frequency power-law index $\alpha_{hf}$ is around zero. The flattened spectra may indicate that the dynamics at the high frequencies are largely uncorrelated in time. Such behaviour could arise from turbulence development towards sufficiently small scales where correlations are rapidly destroyed \citep[e.g.,][]{2008PhRvL.100f5004H}. Alternatively, it may reflect a superposition of small-scale independent nanoflares \citep[e.g.,][]{2015RSPTA.37340256K,2015ApJ...798....1I}. In addition, finite plasma response times in thermal conduction and radiative cooling \citep[e.g.,][]{1995ApJ...439.1034C} could effectively smooth high-frequency intensity fluctuations, thereby suppressing temporal correlations. Furthermore, instrumental noise, which is typically time-independent \citep[e.g.,][]{2012SoPh..275...41B}, may also contribute to the apparent flattening of the power spectrum at high frequencies. These possible mechanisms could operate simultaneously, producing the observed stochastic signal.

As mentioned above, the reversal between $\alpha_\mathrm{lf}$ and $\alpha_\mathrm{hf}$, represented by $\min(\alpha_\mathrm{lf}-\alpha_\mathrm{hf})<-1$, reflects a sharp variation of the EUV intensity, caused by various small-scale transients in an AR. This is confirmed by tracing its location and occurrence time back to the EUV image sequence. The $\min(\alpha_\mathrm{lf}-\alpha_\mathrm{hf})<-1$ corresponds to brightenings/jets lasting for several minutes. The spatial structure of corresponding brightenings is of a dotted shape, Y-shape or elongated shape, or propagates along threads, which are typical signatures of magnetic reconnection \citep[e.g.,][]{2019LRSP...16....3T, 2025arXiv250507520D}. Additionally, similar intensity power spectra that are flat in low frequencies but are steep at high frequencies have been predicted in the case of a cluster of flares or nanoflares \citep[e.g.,][]{1991SoPh..133..357H}.

\begin{figure}
    \centering
    \includegraphics[width=0.9\linewidth]{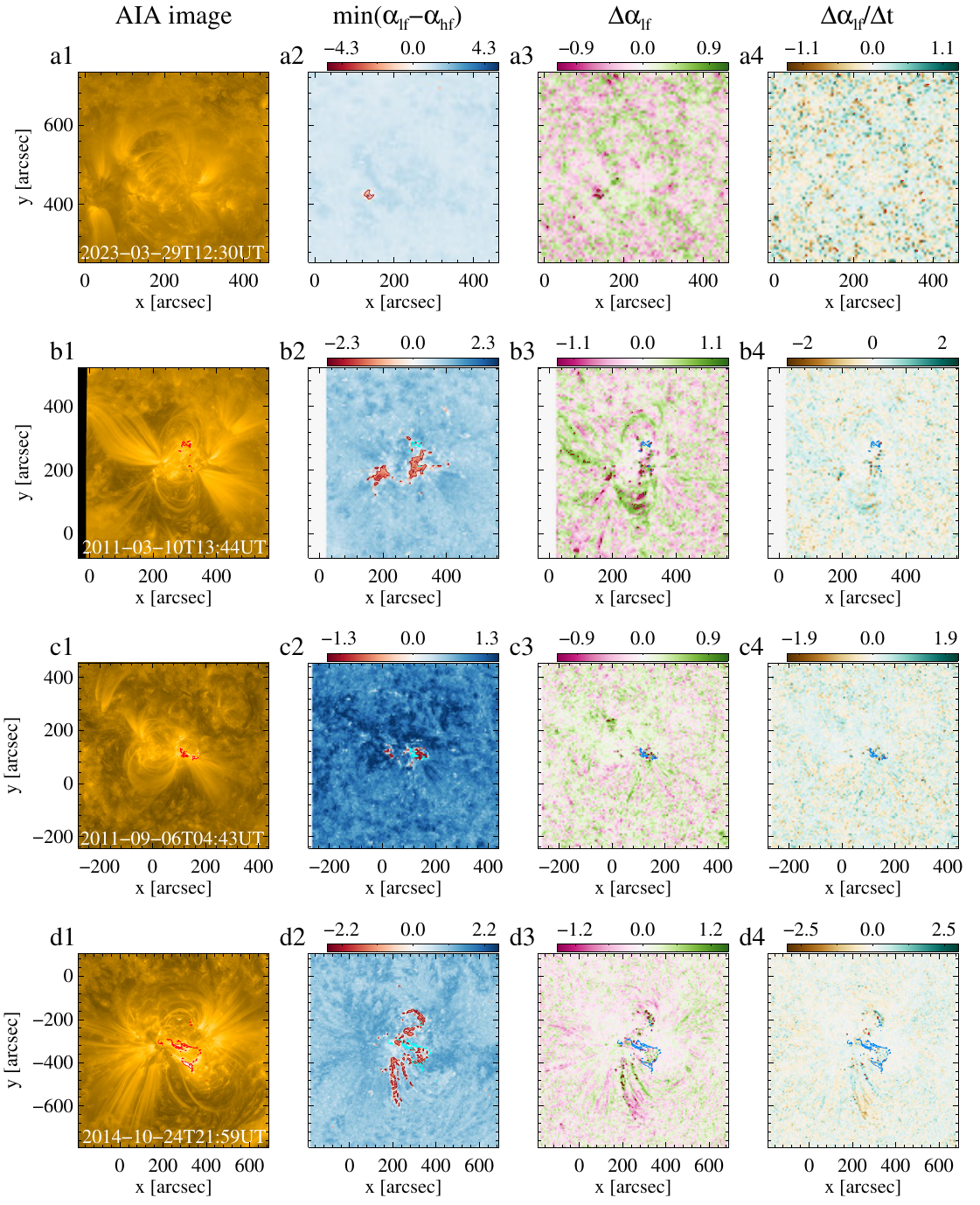}
    \caption{Three potential proxies as flare precursors in four ARs of interest, with the first row for a non-flaring AR and the second to fourth row for a representative C/M/X class flaring AR, respectively. The first column: AIA images showing the ARs of studied, overlaid with flaring structures outlined by the red contours. These structures are duplicated (in blue) in the remaining panels in the same row to show the flare site. The second to forth column: map of $\min(\alpha_\mathrm{lf}-\alpha_\mathrm{hf})$ showing the maximum negative difference between $\alpha_\mathrm{lf}$ and $\alpha_\mathrm{hf}$, map of $\Delta \alpha_\mathrm{lf}$ showing the maximum time difference of $\alpha_\mathrm{lf}$, and map of $\Delta\alpha_\mathrm{lf}/\Delta t$ showing the growth rate of $\alpha_\mathrm{lf}$ in the unit of per hour. Contours in red in the 2nd, 3rd and 4th column indicate $\min(\alpha_\mathrm{lf}-\alpha_\mathrm{hf})=-1$, $\pm0.6\max(\Delta \alpha_\mathrm{lf})$, $\pm0.7 \max(\Delta\alpha_\mathrm{lf}/\Delta t)$, respectively.}
    \label{fig:summary}
\end{figure}

\begin{figure}
    \centering
    \includegraphics[width=0.75\linewidth]{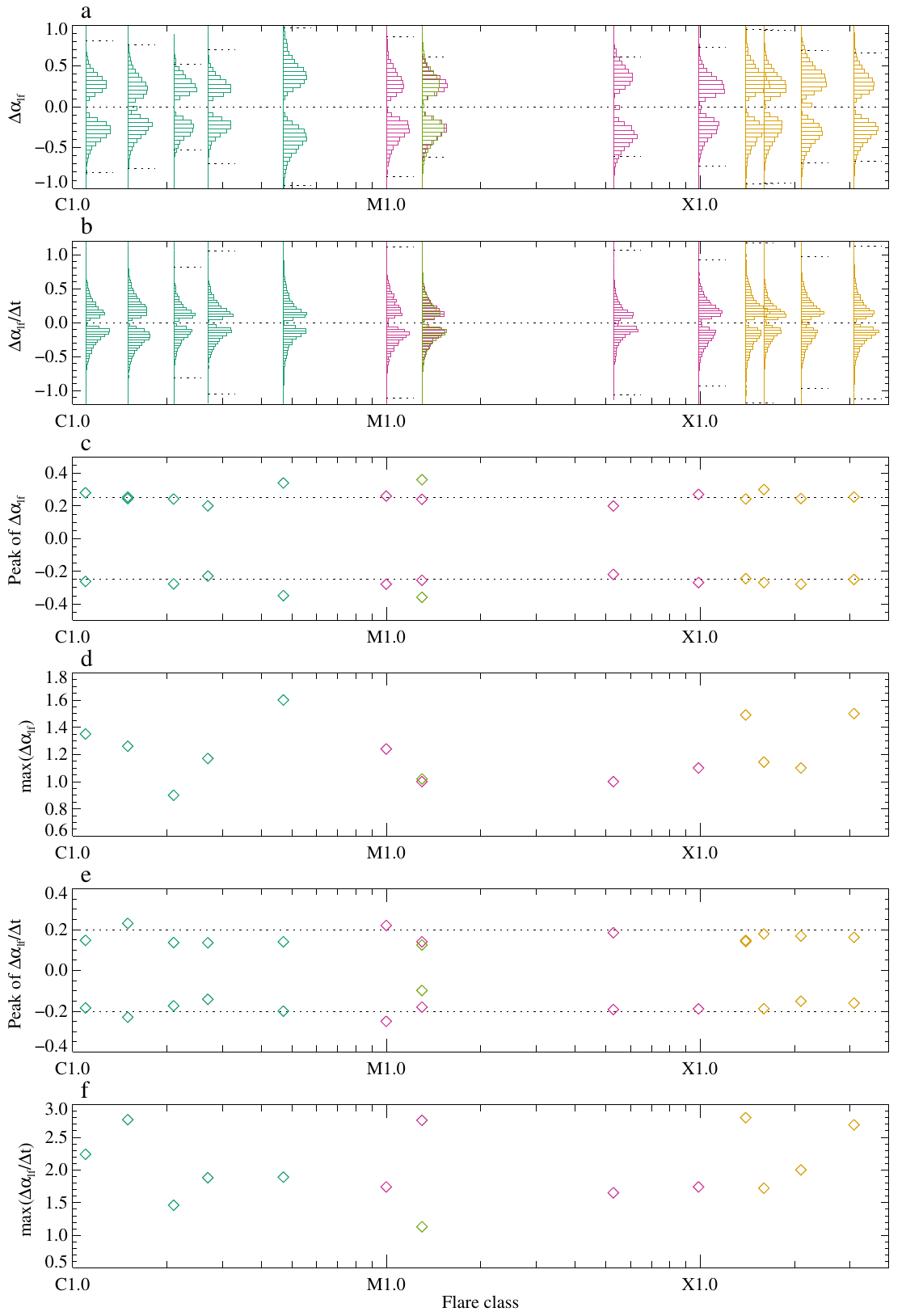}
    \caption{
    The distribution of $\Delta \alpha_\mathrm{lf}$ in relative frequency (a), $\Delta\alpha_\mathrm{lf}/\Delta t$ in relative frequency (b), the peak value of $\Delta \alpha_\mathrm{lf}$ (c), the maximum value of $\Delta \alpha_\mathrm{lf}$ (d), the peak value $\Delta\alpha_\mathrm{lf}/\Delta t$ (e), and the maximum $\Delta\alpha_\mathrm{lf}/\Delta t$ (f) as a function of flare class. The green one in M-class is 2023-03-29 AR 13262. The horizontal lines in (a) indicate the $\pm0.6\max(\Delta\alpha_{\rm lf})$ and those in (b) stand for $\pm0.7\max(\Delta\alpha_{\rm lf}/\Delta t)$ for each case. A missing horizontal line means it is beyond the presented range. cf: for a quiet AR 13264, $\max(\Delta \alpha_\mathrm{lf})=1.21$ , $\max(\Delta\alpha_\mathrm{lf}/\Delta t)=1.35$. No correlation between proxies and flare magnitude is found.
    }
    \label{fig:correlation}
\end{figure}

\begin{figure}
    \centering
    \includegraphics[width=\linewidth]{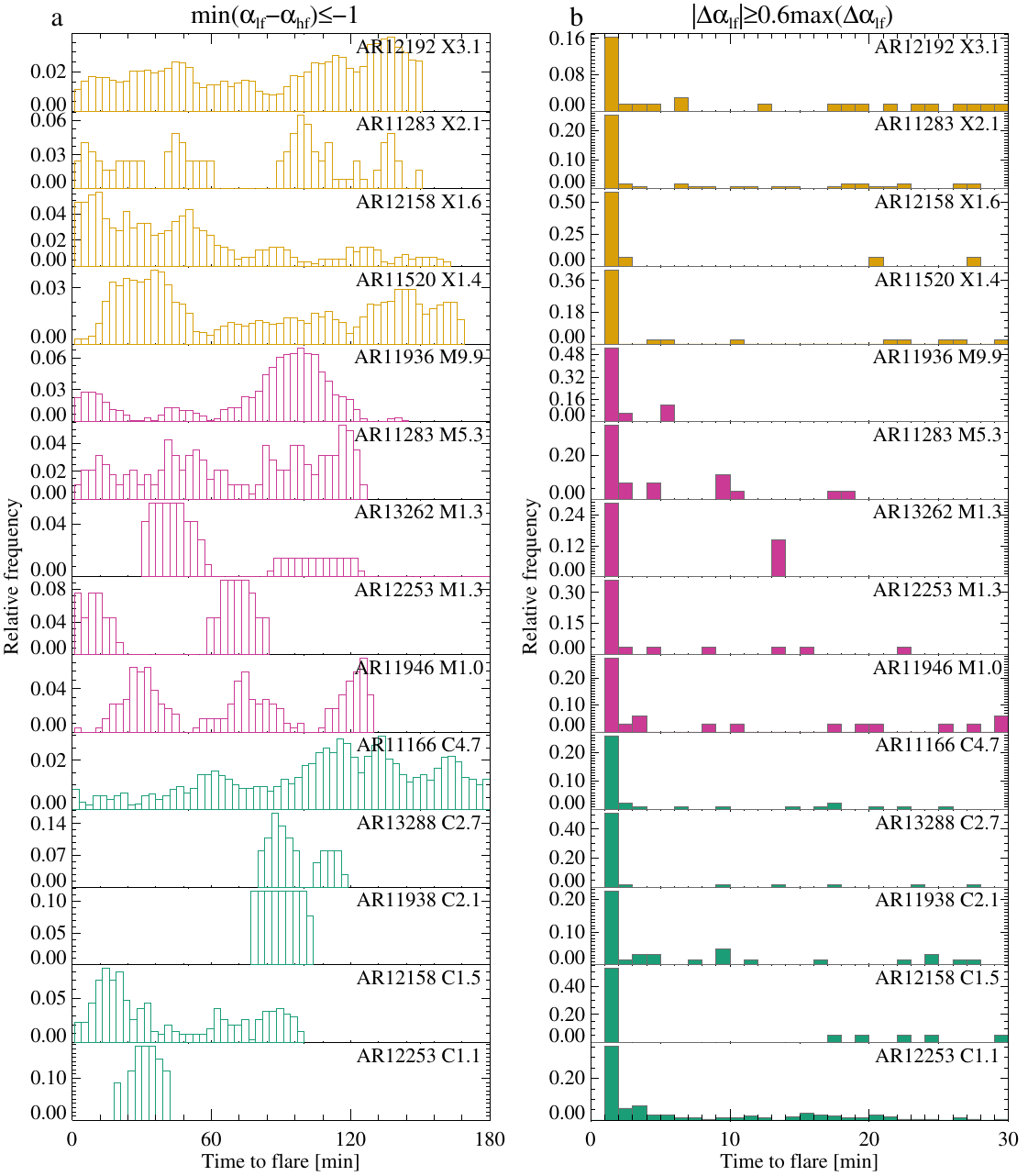}
    \caption{Arrival time to flare time when $\min(\alpha_\mathrm{lf}-\alpha_\mathrm{hf})<-1$ (a) and when $|\Delta\alpha_\mathrm{lf}|$ exceeds $0.6\max(\Delta\alpha_\mathrm{lf})$ occurs (b) in each event of analysis. The three vertical dotted lines indicate 30, 60, and 90~min to flare from left to right. 
    }
    \label{fig:time}
\end{figure}

\begin{figure}
    \centering
    \includegraphics[width=\linewidth]{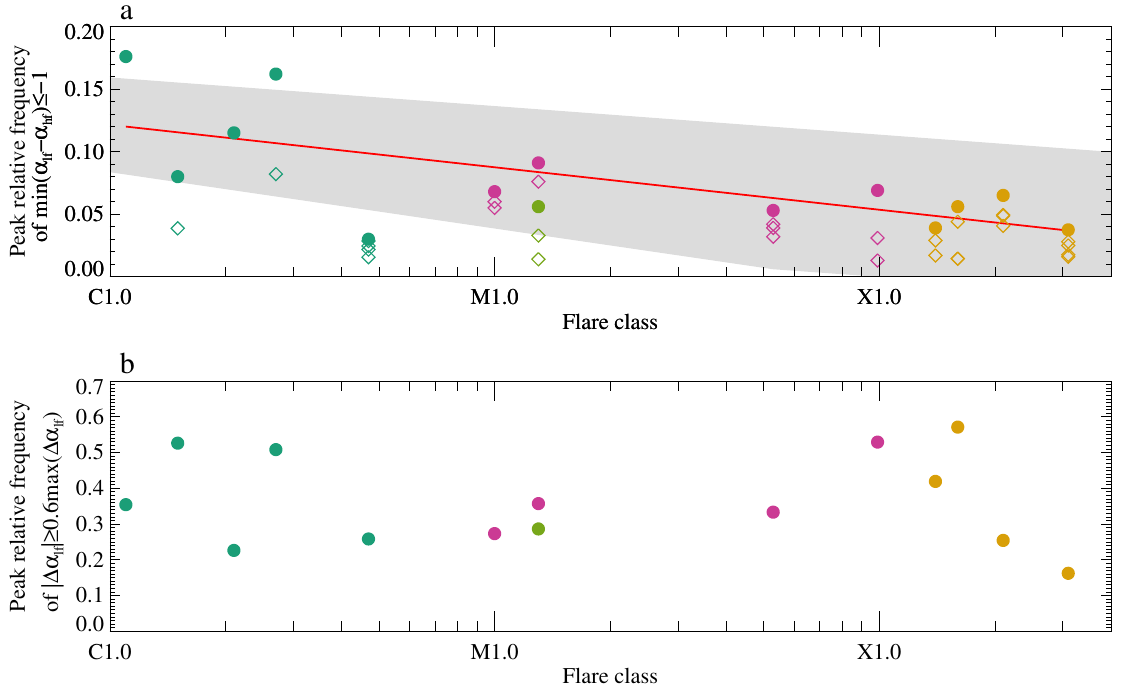}
    \caption{
    Peak relative frequency of the occurrence of $\min(\alpha_{\rm lf}-\alpha_{\rm hf})\leq -1$ (a) and $|\Delta\alpha_\mathrm{lf}|\geq0.6\max(\Delta\alpha_\mathrm{lf})$(b) as a function of the flare class. In panel (a), the values for each peak in the time distribution in Figure~\ref{fig:time}(a) are selected and shown in each analysed case, and the maxima are indicated by the filled symbols. The green symbols in M1.3 correspond to 29th March 2023 event. The red curve shows the best fit to the maximum values for each case (filled symbols), with the fitting function $peak\,frequency=0.15-0.01\log(flare\_class)$, and the associated uncertainty shown as the grey shaded area.
    }
    \label{fig:cc_freq}
\end{figure}

Considering the physical meaning of the increase or decrease in $\alpha_\mathrm{lf}$ prior to flares, the shifting dominance of longer or shorter dynamic timescales suggests the presence of an as-yet unknown process that may be closely linked to the flare ignition mechanism. This interpretation is conceptually analogous to the forecasting of catastrophic failure in structural dynamics, where changes in the frequency content of vibrations are used as precursors to breakdown \citep[see][for a recent review]{2021MSSP..14707077A}.
Indeed, a high value of $|\Delta\alpha_\mathrm{lf}|$ could be a response to the effective magnetic energy build-up process accompanied by short-scale dynamic processes, while a low value of $|\Delta\alpha_\mathrm{lf}|$ represents a quasi-steady state of the underlying magnetic activity.
Moreover, the increase in $\alpha_\mathrm{lf}$ in time means that the Fourier energy decreases faster with increasing frequency, and more energy goes to low frequencies. This may indicate a process with a timescale of tens of minutes rather than seconds for free magnetic energy accumulation. 
In some cases (for example, Figures~\ref{fig:summary}(b3) and (d3)), the increase and decrease in $\alpha_\mathrm{lf}$ happen nearly simultaneously in the same active region structures, which may indicate the local nature of this energy exchange process. Although the underlying physical cause of the continuous increase or decrease of $\alpha_\mathrm{lf}$ before flares remains unknown, its location and timing suggest a strong correlation with flare onset, indicating its potential value for flare prediction.


The locations where our proxies exceed the thresholds do not always coincide precisely with the flare site. This is attributed to the fact that the flare trigger could be elsewhere rather than near the flaring structure. 
For example, \citet{2022Ge&Ae..62..356Z} reported that the sources of preflare QPPs are not necessarily located within flaring ARs and, in some cases, appear in spatially separated ARs.
Moreover, extended cusp regions have been observed in solar flares \citep{2022Natur.606..674F}, with the location of reconnection shifted away from the flaring magnetic arcades \citep{2011SoPh..273..363B, 2021SoPh..296..185K}, so that electrons may get accelerated not only at the site of primary energy release but also due to local dynamic interactions of the magnetic plasma structures residing near the reconnection site.
Such findings challenge the standard flare model and suggest that factors such as magnetic elasticity and convection may play a role in flare initiation.

There are limitations for analysing spectra spanning less than a decade of frequency. This can limit the model validation and potentially affect the accuracy of the parameter estimation \citep[e.g.,][]{2009SIAMR..51..661C, 2016SoPh..291.1561D}. The former could be addressed by performing model comparison by a Bayesian approach \citep{2021ApJS..252...11A, 2022FrASS...926947A}, which is out of the scope in this work. The narrow high-frequency range is an inherent limitation of the instrumentation, as the highest frequency in the spectra is constrained by the temporal resolution of the imaging instrument. However, we note that high-cadence observations from instruments like the Extreme Ultraviolet Imager (with a time cadence of 2~s) onboard Solar Orbiter exhibit similar white-noise-like patterns \citep[][]{2023A&A...678A.188G}, suggesting our model captures the underlying dynamics well. Future observational tools with higher temporal resolutions, such as the SKA \citep[10~ms to 100~ms;][]{2009IEEEP..97.1482D,2019AdSpR..63.1404N}, Solar-C \citep[0.5~s;][]{2019SPIE11118E..07S}, MUSE \citep[0.5~s;][]{2022ApJ...926...52D} or SPARK \citep[0.1~s,][]{2023Aeros..10.1034R}, will allow for extending the frequency ranges and allow for more robust model comparisons.

This proof-of-concept study offers initial evidence that temporal changes in the power-law indices of Fourier power spectra of coronal EUV emission intensity may be associated with flare onset, based on a limited dataset of 14 events. Future work should incorporate a larger sample of flare events to evaluate the prevalence and reliability of pre-flare variations in $\alpha_\mathrm{lf}$ and $\alpha_\mathrm{hf}$, and diagnose their predictive value. Other potential applications of Fourier power spectrum analysis include investigation on whether the temporal evolution of $\alpha_\mathrm{lf}$ and $\alpha_\mathrm{hf}$ is synchronised across different atmospheric heights. For example, this could shed light on processes such as the upward propagation of magnetoacoustic waves and their transition into turbulence \citep[see, e.g.,][]{2018ApJ...861...33K, 2018ApJ...856...73C}, and could form the basis of a future follow-up study.




\begin{acknowledgments}
\textbf{Acknowledgments}\\
This work is supported by the Latvian Science Council Grant lzp-2024/1-0023 and the DynaSun project under the Horizon Europe programme of the European Union under grant agreement (no. 101131534).
S.Z. acknowledges support by an FWO (Fonds voor Wetenschappelijk Onderzoek -- Vlaanderen) postdoctoral fellowship (1203225N). 
D.Y.K. thanks the UKRI Stephen Hawking Fellowship EP/Z535473/1.
V.M.N. is supported by ERC grant 101201424 (ACDCSUN).
V.M.N. and D.Y.K. acknowledge funding from UK Research and Innovation under the UK government the Horizon Europe funding guarantee EP/Y037456/1. 
For the purpose of open access, the authors have applied for a Creative Commons Attribution (CC BY) license to any Author Accepted Manuscript version arising.\\
\\
\textbf{Data Availability}\\
SDO/AIA images are available in JSOC (\url{http://jsoc.stanford.edu/}). Proxy map sequences for all analysed cases are available at Zenodo: \url{10.5281/zenodo.16100847}. In this repository, each video represents an analysed case, named by the corresponding event and flare magnitude. Each video shows four panels in the same format as each row of Figure 4: EUV image and three proxy maps, overplotted with the contours of thresholds as described in the text.
\end{acknowledgments}

%

\vspace{5mm}
\facilities{SDO/AIA}


\software{SSWIDL \citep{1998SoPh..182..497F}, Python}





\bibliography{sample631}{}
\bibliographystyle{aasjournal}

\end{document}